\documentstyle[psfig]{mn}

\footnotesize
\newdimen\digitwidth    
\setbox0=\hbox{\rm0}
\digitwidth=\wd0
\catcode`!=\active
\def!{\kern\digitwidth}
\normalsize

\title[A disrupted binary pulsar?]
{PSR~J0609+2130: A disrupted binary pulsar?}
\author[D.~R. Lorimer et al.]
{
D.~R. Lorimer$^1$,\thanks{Email: drl@jb.man.ac.uk}
M.~A. McLaughlin$^1$,
Z. Arzoumanian$^2$,
K.~M. Xilouris$^3$,
\newauthor
J.~M. Cordes$^4$,
A.~N. Lommen$^5$,
A.~S. Fruchter$^6$,
A.~M. Chandler$^7$,
D.~C. Backer$^8$
\\
$^1$ University of Manchester,
Jodrell Bank Observatory, Macclesfield, Cheshire, SK11~9DL, UK\\
$^2$ USRA/LHEA, NASA Goddard Space Flight 
Center, Code 662, Greenbelt, MD 20771, USA\\
$^3$ University of Virginia, Department of Astronomy, PO Box 3818,
Charlottesville, VA 22903, USA\\
$^4$ Astronomy Department, Cornell University, Ithaca, NY 14853, USA\\
$^5$ Department of Physics \& Astronomy, Franklin \& Marshall College,
PO Box 3003, Lancaster, PA 17604, USA\\
$^6$ Space Telescope Science Institue, 3700 San Martin Drive, Baltimore, 
MD 21218, USA\\
$^7$ Space Radiation Laboratory, California Institute of Technology, MS 220-47, Pasadena, CA 91125, USA\\
$^8$ Astronomy Department, University of California, Berkeley, CA 94720, USA}

\date{Accepted for publication in MNRAS 11/11/03}
\begin{document}

\maketitle
\newcommand{\setthebls}{
}

\setthebls

\begin{abstract} 
We report the discovery and initial timing observations of a 55.7-ms
pulsar, J0609+2130, found during a 430-MHz drift-scan survey with the
Arecibo radio telescope. With a spin-down rate of $3.1 \times
10^{-19}$ s s$^{-1}$ and an inferred surface dipole magnetic field of
only $4.2\times10^{9}$~G, J0609+2130 has very similar spin parameters
to the isolated pulsar J2235+1506 found by Camilo, Nice \& Taylor
(1993). While the origin of these weakly magnetized isolated neutron stars
is not fully understood, one intriguing possibility is that they are the
remains of high-mass X-ray binary systems which were disrupted
by the supernova explosion of the secondary star.
\end{abstract}


\section{Introduction}
\label{sec:intro}

Radio pulsars are categorized as either ``normal'' or ``recycled''
objects.  Normal pulsars are observationally more numerous and are
predominantly young isolated objects with spin periods in the range 30
ms $< P < 8$ s, strong inferred surface dipole magnetic field
strengths ($B\sim 10^{12}$ G) and characteristic ages in the range
$10^3<\tau< 10^7$ yr. Recycled pulsars have shorter spin
periods (1.5 ms $<P<60$ ms), weaker magnetic fields ($10^8<B< 10^9$
G), larger ages ($10^8 < \tau < 10^{10}$ yr) and are usually members
of binary systems with either white dwarf or neutron star companions.

The normal pulsars are thought to be born in supernovae and spin down
due to magnetic dipole braking \nocite{go70} (see e.g.~Gunn \&
Ostriker 1970).  The favoured theory for the origin of recycled
pulsars involves an old neutron star that is spun up (recycled)
through the accretion of matter from a binary companion which
overflows its Roche lobe \nocite{bk74} (Bisnovatyi-Kogan \& Komberg
1974).  During the spin-up phase, the system is expected to be visible
as an X-ray binary. The duration of the X-ray phase and amount of spin
up depends upon the mass of the companion star. For those companions
massive enough to explode as a supernova, the X-ray lifetime is
relatively short ($10^{6-7}$ yr) and the most likely outcome is the
disruption of the binary. Those systems fortunate enough to survive
the explosion are the double neutron star (hereafter DNS) binaries
(e.g., the original binary pulsar B1913+16). For less massive
companions, where the period of spin up is longer ($10^8$ yr), the
subsequent collapse leaves a white dwarf star in orbit around a
rapidly spinning millisecond pulsar. For a detailed review of these
evolutionary scenarios, \nocite{bv91} see Bhattacharya \& van den
Heuvel (1991) and references therein.

In order to understand the population of recycled pulsars in detail, a
statistically significant sample is required. Because of their short
spin periods and, often, binary orbits, detecting these elusive
objects is a computationally intensive process requiring
state-of-the-art data acquisition systems, significant data storage
and computational power for post processing.  
In 1994 the Penn State Pulsar Machine
(PSPM), a new analogue filterbank spectrometer (Cadwell
1997), \nocite{cad97a} was installed at the 305-m Arecibo telescope
and has been in regular use as a pulsar search and timing instrument
ever since. In March 2002, we began using part of a high-speed
computer cluster to process various sets of PSPM 
data taken during the latter stages
of the Arecibo upgrade (1996--8). To date, we have discovered a dozen
new pulsars and re-detected over 40 previously known ones. Preliminary
details of the search were given by \nocite{mla+03} McLaughlin et al.~(2003). 

Here we report on an isolated 55.7-ms pulsar J0609+2130, one of the 
first new discoveries from our searches. While the spin parameters
of the new pulsar are similar to other recycled pulsars, the lack of a
binary companion poses interesting questions as to its origin.
The layout of the rest of this paper is as follows: in \S \ref{sec:obs} we
briefly describe the survey observations, data acquisition system and
analysis pipelines. The results of the analysis and subsequent timing
observations of J0609+2130 are given in \S
\ref{sec:res}. Finally, in \S \ref{sec:disc}, we discuss the new
pulsar's properties in the context of its likely origin and evolution.

\section{Survey observations and analysis}
\label{sec:obs}

The data presented here were taken with the Arecibo radio telescope
during a 9.5-hr maintenance period
in October 1998 when upgrade operations required that the telescope was
parked at $35.7^{\circ}$ azimuth.  The 430-MHz line-feed system,
traditionally used for pulsar observations, was not available for use
on this occasion. We therefore took advantage of the newly commissioned
430-MHz receiver in the Gregorian dome which was in a fixed position at
$3.9^{\circ}$ from the zenith.  Although the 430-MHz Gregorian
receiver\footnote{see \verb+http://www.naic.edu/\~\,astro/RXstatus+ for
full details of currently available Arecibo receiving systems}
illuminates less of the dish than the line-feed so that the
forward gain is lower (11 K/Jy cf.~18 K/Jy for the line feed), the lower 
system temperature (45 K cf.~120 K) and reduced levels
of interference make it a very attractive receiver for pulsar searching.
In this setup, sources at declination $\delta$ drift through
the 11-arcmin primary beam in $\sim 44/\cos \delta$ s. The total
area covered was $\sim 22$ deg$^2$ along an 11-arcmin slice through
the Galactic anti-centre in the right 
ascension range $1.5^{\rm h} < \alpha < 10^{\rm h}$ centred at 
$\delta \simeq 21.5^{\circ}$.

The incoming signals from the telescope were passed to the PSPM which
summed the two independent polarizations before 4-bit sampling the
band into $128\times60$-kHz  channels every 80 $\mu$s. For this
choice of sampling interval, a $2^{19}$-pt Fourier transform is
well matched to the 44-s transit time of sources through the 430-MHz beam. 
The resulting data were
then written directly to magnetic tape for offline analysis. 

Searches for periodic and transient radio signals in the data were
carried out using COBRA, a 180-processor Beowulf cluster at Jodrell
Bank Observatory. 
Deferring a more complete discussion for a future paper, we now
briefly outline the main procedures.  To correct for the dispersive
effects of the interstellar medium, the raw PSPM data were dedispersed
at 392 different trial dispersion measures (DMs) in the range $0 \leq$
DM $\leq 491.2$ cm$^{-3}$ pc using freely available analysis tools
\nocite{lor01b} (Lorimer 2001). The resulting dedispersed time series
were then passed to two different analysis pipelines.  In the first,
the data were Fourier transformed and the resulting amplitude spectra
were searched for significant features indicating the presence of a
periodic signal. To increase the sensitivity to narrow duty cycle
pulses, spectra summing 2, 4, 8 and 16 harmonics were also
searched. Candidates with signal-to-noise (S/N) ratios above 8 were then
analysed in the time and radio frequency domain to produce diagnostic
plots which were saved for later
visual inspection. Further details can be found in \nocite{lkm+00}
Lorimer et al.~(2000). 

In the second analysis pipeline, the
dedispersed data were searched in the time domain for pulses
with S/N $>5$ using
software developed by Cordes \& McLaughlin \nocite{cm03} (2003).
This search was aimed at
finding individual and giant pulses from some pulsars where the
periodic signal may be below the threshold of the Fourier analysis
(see e.g.~Nice \nocite{nic99} 1999).  To maximize sensitivity over a
variety of pulse widths, each time series was smoothed by 
up to 64 adjacent samples. 

Given the measured system
equivalent flux density of the 430-MHz Gregorian receiver (3.3 Jy), we
estimate the sensitivity of the Fourier analysis to be
$\sim 0.5$ mJy to long-period pulsars observed away
from the Galactic plane. For millisecond pulsars, with typical periods
of 5 ms and pulse duty cycles of order 60\%, we estimate the sensitivity to
be closer to 3 mJy. The detection threshold of the single-pulse search
is about 0.5 Jy to the narrowest pulses. 
Full details of the search sensitivity will be presented elsewhere. 

\section{Results and timing observations}
\label{sec:res}

Two pulsar-like signals were detected in the periodicity search. One
of these was the 33-ms Crab pulsar, B0531+21 (also seen in the single-pulse
analysis) detected in several adjacent beams with S/N $\sim 18$. The
remaining candidate, a 55.7-ms signal with DM $\sim 40$ and S/N $\sim
9$, was re-observed and confirmed as the new pulsar J0609+2130 in April 2002.
Only one other previously known pulsar, B0525+21, lies within the 
region covered by this search. This was not detected in either the
periodicity or the single-pulse search. A closer scrutiny of the
raw search data revealed no detectable signal. Since the data were not
obviously affected by radio-frequency interference, two possible
effects are responsible for the non-detection: either (a)
since this pulsar is known to null for approximately 25\% of the time
(Biggs 1990) \nocite{big90b} it is quite likely that the survey
observations caught this pulsar in a null state; or (b) the pulsar's
flux density was significantly reduced by diffractive
interstellar scintillation. Given the normally high mean 
flux density of this pulsar (57 mJy at 408 MHz; Lorimer et
al.~1995), \nocite{lylg95} nulling seems the most likely
explanation for the non-detection.

To determine the detailed spin and astrometric parameters
of J0609+2130, we have been carrying out regular 430-MHz timing observations
since June 2002. These observations use the PSPM in timing
mode, where the incoming signals are folded modulo the period predicted
from an ephemeris initially derived from the confirmation observation.
In timing mode, folded 1024-bin pulse profiles for each of the 128 frequency 
channels of the PSPM are written to disk every 180 s. We initially collected
at least three of these 180-s observations per epoch in order to
enable high-precision estimates of the pulse period. The timing analysis 
proceeded by forming the dedispersed profile across the band of 
the PSPM for each 180-s observation and estimating the pulse phase
by cross correlating each profile with a high S/N template
using procedures identical to those described in detail by \nocite{lcx02}
Lorimer, Camilo \& Xilouris (2002). 

\begin{table}
\caption{\label{tab:timing}Observed and derived parameters for PSR~J0609+2130}

\begin{tabular}{ll}
\hline
\hline
\noalign{\smallskip}
Parameter & Value \\
\noalign{\smallskip}
\hline
\noalign{\smallskip}
Right ascension (h:m:s) (J2000) & 06:09:58.883(1) \\
Declination (deg:m:s) (J2000) & 21:30:02(1) \\
Spin period, $P$ (ms) &  55.6980139253(2) \\
Epoch of period (MJD) &  52575.0 \\
Period derivative, $\dot{P}$ ($\times 10^{-19}$ s s$^{-1}$) & 3.1(6) \\
Dispersion measure, DM (cm$^{-3}$ pc) & 38.77(5) \\
Mean flux density at 430 MHz, $S_{430}$ (mJy) & 0.8(1)\\
Galactic longitude, $l$ (deg) & 189.192 \\
Galactic latitude, $b$ (deg)  & 1.04\\
\hline
Characteristic age, $\tau$ (Gyr) & 2.8\\
Magnetic field strength, $B$ ($10^9$ G) & 4.2\\
Distance, $d$ (kpc) & 1.2 \\
430-MHz radio luminosity, $L_{430}$ (mJy kpc$^2$) & $\sim$1.15\\
\noalign{\smallskip}
\hline
\end{tabular}

The numbers in parentheses represent 1-$\sigma$ uncertainties
in the least significant digit quoted and are twice the formal
fit estimates from TEMPO. The characteristic age
and magnetic field are calculated as described in \S \ref{sec:disc}.
The distance is derived using the Galactic electron density model
of Cordes \& Lazio (2003).
\end{table}

\begin{figure}
\psfig{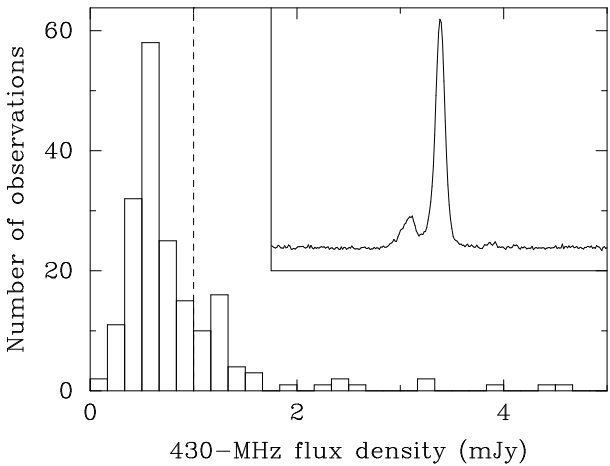}
\caption{
Distribution of 430-MHz flux densities for PSR~J0609+2130 based
on individual 180-s observations.
The dashed line shows the detection threshold for this pulsar
in our original search-mode observation. Inset:
Integrated 430-MHz profile showing 360 degrees of
rotational phase. The profile was
produced by phase-aligning and summing all the 430-MHz
detections. The equivalent integration time and
effective time resolution are 9.8 hr and 256 $\mu$s respectively.
\label{fig:430prof+flux}
}
\end{figure}

The topocentric pulse arrival times were fitted to a simple timing model
involving spin and astrometric parameters using the 
TEMPO\footnote{\verb+http://pulsar.princeton.edu/tempo+} software package.
Once a preliminary
phase-connected timing solution was obtained, a better template was formed by
phase-aligning the pulse profiles according to the best-fit model.
The final template profile is shown in Fig.~\ref{fig:430prof+flux}.  As
part of this refinement procedure, individual 180-s observations on a
given day were phase aligned and averaged together to produce one
pulse arrival time per observing session.  The estimate of DM from
the search analysis was improved by analysing PSPM
observations with the 327-MHz and 610-MHz Gregorian receivers on
February 17, 2003, and using {\sc TEMPO} to fit for the dispersion delay
between the frequency bands.  The best-fit timing model
for 30 arrival times spanning the MJD range 52427--52796 results
in featureless timing model residuals with  a post-fit RMS of 44 $\mu$s.
The resulting parameters are given in Table \ref{tab:timing}. 

\section{Discussion}
\label{sec:disc}

\subsection{Basic properties of J0609+2130}

PSR~J0609+2130 is a very weak radio source.  Using the calibration
procedures described by Lorimer, Camilo \& Xilouris (2002), we
find the mean 430-MHz flux density of the pulsar $S_{430}$ to be
only $0.8\pm0.1$ mJy at 430 MHz. Given the distance estimate $d \simeq 1.2$ kpc
from the new electron density model of Cordes \& Lazio \nocite{cl03}
(2003), the 430-MHz radio luminosity $L = S d^2 \sim 1.15$ mJy
kpc$^2$. Only 10 out of the 612 pulsars currently in the online
catalogue\footnote{\verb+http://www.atnf.csiro.au/research/pulsar/psrcat+} 
with measured luminosities are fainter sources than PSR~J0609+2130.

The pulsar's flux density is clearly affected by interstellar scintillation.
We observe variations between 0.1 and 5 mJy as shown in 
Fig.~\ref{fig:430prof+flux}. Indeed, for about 75\% of the observations
we have taken, the flux density is below our nominal
survey threshold of about 1 mJy for this pulsar. Although our
observations at other frequencies are not yet numerous enough to
characterize the radio spectrum of J0609+2130 with any certainty, our
preliminary flux estimate of 2 mJy at 327-MHz implies that it
is a steep-spectrum source that is most readily detectable in
low-frequency surveys. 

Our timing measurements of PSR~J0609+2130 reveal a period derivative 
$\dot{P} = 3\times 10^{-19}$. PSR~J0609+2130 therefore appears to be a 
an old neutron star with characteristic age $\tau = P / (2\dot{P}) = 2.8$
Gyr and a relatively weak surface dipolar magnetic field $B = 3.2
\times 10^{19} \sqrt{P \dot{P}} = 4.2 \times 10^9$ G.  
Of all the sources in the current pulsar catalogue, the one which has 
properties most similar to J0609+2130 is PSR~J2235+1506, a solitary 57.9-ms
pulsar with $\dot{P}=1.7 \times 10^{-19}$ \nocite{cnt93,cnt96}
(Camilo, Nice \& Taylor 1993, 1996). The positions of both these pulsars on
the $B-P$ diagram are shown in Fig.~\ref{fig:bp}. As can
be seen, J0609+2130 and J2235+1506 are the only two solitary pulsars
in this part of the diagram. We shall return to the positions
in the $B-P$ plane later in the discussion.

\subsection{The Galactic population and evolution of weakly magnetized solitary pulsars}

Based on the close proximity and low radio luminosities of J0609+2130
and J2235+1506 we expect pulsars like them to be quite common in the
Galaxy. Indeed, Kalogera \& Lorimer (2000) \nocite{kl00} estimated 
the Galactic population
of pulsars like J2235+1506 to be $\sim 5000/f$, where $f$ is the
fraction of $4\pi$ sr covered by the radio beam. For a beaming
fraction of 30\% (i.e.~$f=0.3$) the Galactic population of these objects is
likely to be of order 15,000. 

What is the origin of weakly magnetized isolated pulsars like J0609+2130
and J2235+1506? One possibility is that they are simply part of
the same population as the normal pulsars. To investigate 
this idea quantitatively, we can take the best-fitting
initial magnetic field distribution in a recent population study of
isolated pulsars by Arzoumanian, \nocite{acc02} Chernoff \& Cordes
(2002). For their distribution, only about one in the expected Galactic
population of $10^9$ neutron stars is born with $B<10^{10}$ G,
clearly inconsistent with the population estimates made above.
Unless there is a weak-field component to the neutron star
magnetic field distribution that has been overlooked by population
studies so far, a more efficient means of forming these
weakly-magnetized neutron stars is required.

A pulsar born with a stronger magnetic field could, in principle,
evolve to resemble PSR~J0609+2130 through long-term decay of the
magnetic field. While the existence of field decay for the
pulsar population at large remains controversial (see e.g.~Bhattacharya 
et al.~1992), \nocite{bwhv92} available
evidence suggests that, for objects comparable in age and
magnetic-field strength to J0609+2130, any decay must act on
extremely long timescales. Where such pulsars are found in
binary systems, studies of their white-dwarf
companions provide an independent age estimate through cooling
models, affirming the long-lived nature of the pulsars
and constraining the existence of field decay (Kulkarni 1986).
\nocite{kul86} Of course, J0609+2130 is an isolated
pulsar, but it seems unlikely that non-binarity would somehow
encourage rapid field decay.

\begin{figure}
\psfig{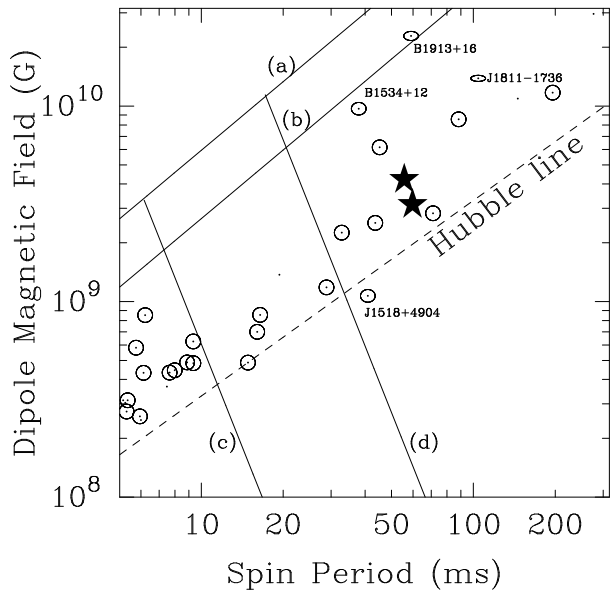}
\caption{Part of the
magnetic field versus period ($B-P$) diagram highlighting the 
unique positions of
the solitary pulsars J0609+2130 and J2235+1506 (starred
symbols). Pulsars with surrounding ellipses are members of binary
systems, with the eccentricity of the ellipse representing the orbital
eccentricity.  Known DNS binary systems are labelled
appropriately. The ``Hubble line'' is the locus of points of constant
characteristic age equal to a Hubble time (assumed to be 15 Gyr).  The
four lines show limiting spin-up values for Alfv\'{e}n accretion
at the Eddington limit and one fifth of Eddington (labelled (a) and (b)
respectively), and terminated accretion after a donor star
life time of $10^6$ yr for the same accretion rates (labelled (c) and
(d)). Pulsars in globular clusters and
the Magellanic clouds have been excluded.
\label{fig:bp}}
\end{figure}

Camilo, Nice \& Taylor (1993) suggested that J2235+1506 might be
the remains of a high-mass binary system that disrupted at the time
of the second supernova explosion. In this scenario, PSRs J0609+2130
and J2235+1506 are the first-born neutron stars in their respective binaries
and initially spin down as regular neutron stars. During Roche-lobe 
overflow from the subsequently evolved secondary star, mass accretion
reduces the magnetic field (see e.g.~Romani \nocite{rom90} 1990),
and also spins up the neutron star in the process. If the secondary
is sufficiently massive to undergo a supernova explosion, the most
likely outcome is disruption of the binary, releasing a newly-born
neutron star and a recycled pulsar. Those binary systems 
more likely to survive this explosion as DNS binaries are the tightly-bound
systems with short orbital periods. The wider binaries, which are much
less likely to survive the explosion, release recycled
pulsars with spin properties akin to J0609+2130 and J2235+1506
into the Galactic field. We hereafter refer to these
as ``disrupted recycled pulsars'' (DRPs).

Using this evolutionary scenario for DRPs as a working hypothesis,
we now return to the $B-P$ diagram in Fig.~\ref{fig:bp} to investigate
the post-accretion spin periods of DRPs after the accretion phase. In
their analysis of DNS binaries, Arzoumanian, Cordes \& Wasserman 
\nocite{acw99} (1999; hereafter ACW) 
discuss two equilibrium spin period relations
for the spin-up of a neutron star in a binary system. The first
of these is the familiar limit for spherical accretion of matter
in which the limiting spin period is set by the corresponding
Keplerian orbital period at the so-called Alfv\'{e}n radius
(see e.g.~Ghosh \& Lamb \nocite{gl92} 1992). On the
$B-P$ plane, this translates to a relationship of the form
$B \propto P^{5/3}$. The constant of proportionality in this
expression depends on a number of factors such as the mass
accretion rate, the radius and moment of inertia of the
neutron star and the opacity of the accreting material.
The uncertainties in these parameters result in a family
of $B \propto P^{5/3}$ curves.
Two such curves, assuming accretion at
the Eddington rate and at one-fifth of this value,
are plotted on Fig.~\ref{fig:bp} using
the relationship quoted by ACW. For any pulsar,
in the absence of magnetic field decay,
we can estimate the spin period
at the end of the accretion phase by tracing a horizontal
line from the current position on Fig.~\ref{fig:bp} to the spin-up line.
Under these assumptions, we infer that the intial spin
period of J0609+2130 and J2235+1506 would have been in the
range 10--20 ms.

ACW pointed out that, for high-mass binary systems, the
above picture is oversimplified since
the lifetime of the donor star will be much less than
the accretion time scale at which the Alfv\'{e}n limit
is reached. They show that this condition produces 
a second family of limiting period curves which,
on the $B-P$ plane, take the form $B \propto P^{-7/2}$.
As noted by ACW, these curves are much more sensitive
to uncertainties in the accretion parameters. In Fig.~\ref{fig:bp}
we plot two such spin-up lines, again assuming accretion at
the Eddington rate and one-fifth of this limit. It is
clear that, for sub-Eddington accretion, the initial
spin period of these DRPs would be much longer, perhaps
very similar to their currently observed spin periods.

As noted by ACW, since pulsar characteristic ages $\tau =
P/(2\dot{P})$ assume a negligible post-accretion spin period, one consequence
of these accretion scenarios is that DNS binaries are younger 
objects than their characteristic ages suggest. If DRPs are 
formed by a similar process, the same applies to them.  
For example, a more likely
``post-spinup'' age for J0609+2130 based on a limiting spin period of
40 ms would be about half its characteristic age (i.e.~about 1.5 Gyr).

\subsection{Testing the DRP hypothesis}

Is the DRP hypothesis consistent with the observations? One parameter
that can be confronted by both theory and observation is the
``survival probability'' of the binary system, hereafter $\eta$,
defined as the fraction of binary systems that remain bound after the
second supernova explosion. Numerous authors have followed the
orbital evolution of a wide variety of binary systems
containing neutron stars using detailed Monte Carlo simulations
(see e.g.~Lipunov \& Prokhorov \nocite{lp84} 1984; Dewey \& Cordes 
\nocite{dc87} 1987; Fryer \& Kalogera \nocite{fk97} 1997). We
use the relatively up-to-date simulations
of Portegies Zwart \& \nocite{py98} Yungelson
(1998) who include a Maxwellian kick velocity distribution with a mean
of 415 km s$^{-1}$ in their
simulations.  From the results presented in Table 1 of their
paper, for this model, we infer $\eta \sim 4\%$.
Based upon this low survival probability, we
therefore expect DRPs to be much more common in the Galaxy than the
DNS binaries. 

To estimate $\eta$ from an observational perspective, we note that the expected
spin parameters of the recycled pulsar we observe in a DNS binary 
are the same as those of the putative DRPs. We can therefore reasonably
expect the pulsars in DNS binaries to have the same radio lifetimes 
as the DRPs. Hence, if $N_{\rm DNS}$ and $N_{\rm DRP}$ are
the total numbers of DNS binaries and DRPs in the Galaxy, it 
follows that 
\begin{displaymath}
  \eta = \frac{N_{\rm DNS}}{N_{\rm DNS}+N_{\rm DRP}}.
\end{displaymath}
Using the results of a recent population study
of DNS binaries by Kim, Kalogera \& Lorimer (2003), we take $N_{\rm
DNS}$ population to be $\sim 1500$ \nocite{kkl03} (a factor of 2
higher than quoted by Kim, Kalogera \& Lorimer 2003, to crudely
account for the population of DNS binaries that do not coalesce in a
Hubble time). Using the above estimate of $N_{\rm DRP} = 15,000$,
we find $\eta \simeq 1500/(15000+1500) = 9\%$. 

One outstanding issue is the observed numbers of DRPs.  Given the
above estimate for $\eta$, we would expect to see roughly 10 DRPs for
each DNS binary. So far, however, we have identified only two DRPs: J0609+2130
and J2235+1506, compared to four DNS binaries in Fig.~\ref{fig:bp}. 
Three main possibilities which might explain this apparently significant
deficit of DRPs are: (a) DRPs are present in the observed sample, but 
are hard to
identify since they have no binary companion.  Although there are
perhaps a further four DRP-like pulsars known (see Table 1 of Kalogera
\& Lorimer 2000), a more thorough analysis of the sample would be
worthwhile.  (b) The neutron star kick velocity distribution could be
lower than assumed in the population synthesis. This would lead to an
increase in $\eta$ and reduce the discrepancy.  A population synthesis 
dedicated to this question would certainly be useful. (c) The above DRP
population estimates are somewhat crude and are based currently on
J2235+1506 which is a nearby, low-luminosity object. A revision of
these estimates to better account for
small-number statistics is strongly recommended.

In summary, while it seems currently plausible that J0609+2130 and 
J2235+1506 are examples of disrupted X-ray binary systems, there are
a number of open issues linking the two populations. Careful population
studies to properly identify and estimate the DRP population and
binary population syntheses to predict the survival rate are both
strongly recommended. Further observational input should come both
in the substantial numbers of pulsars now being found in the Parkes 
Multibeam survey \nocite{mlc+01} (Manchester et al.~2001), and also
proper motion measurements. For J2235+1506, the implied transverse speed
is $V_t = 100\pm40$ km s$^{-1}$  (Camilo, Nice \& Taylor 1996). 
If J0609+2130 has a similar proper motion, it should be
detectable in our on-going Arecibo timing observations by early 2005.

\section*{Acknowledgments} 
The Arecibo observatory, a facility of the National Astronomy and
Ionosphere Center, is operated by Cornell University in
a co-operative agreement with the National Science Foundation (NSF).
We thank Alex Wolszczan for making the PSPM freely available
for use at Arecibo. Without this superb instrument, the results presented
here would not have been possible. We also thank Paulo Freire for
observing assistance and the referee, Simon Johnston, for many useful comments.
DRL is a University Research
Fellow funded by the Royal Society.  MAM is an NSF MPS-DRF Fellow.

\end{document}